*Title of the report*
# Hunting down the cause of solar magnetism
*Research institution*
**[1]Aalto University, Department of Computer Science, Astroinformatics Group, Finland**
*Principal Investigator*
**Maarit Käpylä**[1,2,3]
*Researchers*
Lucia Duarte[2], Johannes Pekkilä[1], Ameya Prabhu[2], Matthias Rheinhardt[1], Mariangela Viviani[2], Jörn Warnecke[2]
*Project partners*
**[2]Max Planck Institute for Solar System Research, SOLSTAR group, Germany**
**[3]Nordita, Stockholm, Sweden**
*SuperMUC-NG project ID*
**pn98qu**

## Introduction

The research of the groups of Astroinformatics in Aalto University, Finland [1], and SOLSTAR at the Max Planck Institute for Solar System Research in Göttingen, Germany [2], focuses on understanding solar and stellar dynamos. We try to achieve this by combining high-resolution, local and global numerical modelling with long-term observations. This is a challenging task: even with state of the art computational methods and resources, the stellar parameter regime remains unattainable. Therefore, we are just able to simulate "tar-stars", namely models in which diffusivities are raised several orders of magnitude from their real values, to guarantee numerical feasibility and stability. This work is part of the PRACE project Access - Call 20 INTERDYNS, which has been granted 57M core hours on SuperMUC-NG. Our goal is to relax some approximations, in order to simulate more realistic systems, and try to connect the results with theoretical predictions and state-of-the-art observations.

Convective turbulence, together with large scale flows, such as differential rotation ($\Omega$-effect), plays a key role in generating and shaping the magnetic fields observed in the Sun and other stars. The most important turbulent effect is the $\alpha$-effect, describing collective inductive action arising from cyclonic turbulence. The $\alpha$- and $\Omega$-effects are the two prominent generators of large scale stellar/solar magnetic fields. These effects escape observational efforts even with the largest existing and planned observational infrastructures. In theory, these effects are parametrised by transport coefficients which collectively describe the effects of turbulence without having the need to resolve smaller scales, hence the numerical determination of these coefficients which describe them is of utmost importance. For their measurement, we employ the test-field method (TFM). In its standard form, it can be used to measure the turbulent transport coefficients in the limit where a primary magnetic background turbulence vanishes (low resolution regime). If the background magnetic turbulence is present, i.e., in the high resolution regime, the non-linear (NL) form of TFM is necessary. This regime is relevant for the Sun, which most likely exhibits vigorous small-scale dynamo action, generating magnetic background turbulence. Moreover, the TFM allows us to measure the full $\alpha$ tensor, therefore adding a considerable refinement to the standard theory, which models the $\alpha$-effect merely by a scalar quantity.

## Results and Methods

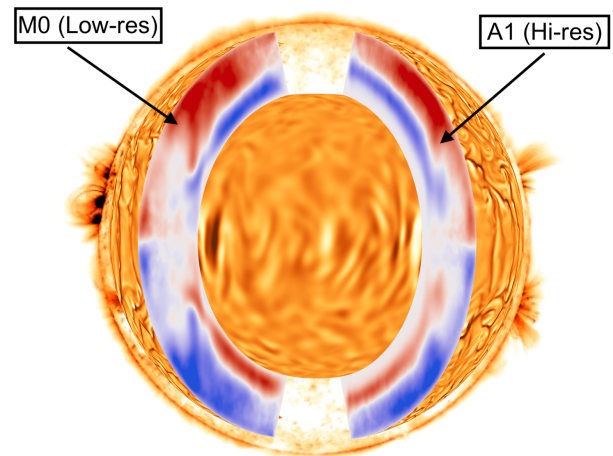

*Figure 1: A volume rendering of global convective dynamo simulations. Colors on the two spherical surfaces represent the radial velocity close to the surface and the bottom of the simulation domain. The meridional cuts show the $\phi\phi$ component of the $\alpha$ tensor for a previous low-resolution run (M0) and the new high-resolution run (A1), red indicating positive, and blue negative values. The background image is from an instrument onboard the Solar Dynamics Observatory.*

We use the Pencil Code [3], a highly modular code, to solve the fully compressible equations of magnetohydrodynamics (MHD). The code employs a sixth-order, central finite differences scheme for spatial discretization and a 3rd-order Runge-Kutta time-integration scheme. The chosen scheme makes the code highly scalable and adaptable. To maintain the magnetic field divergence-free, the code solves for the magnetic vector potential. The output files are written out quite often in order to re-start from crashes or node-failures. The parallelization is implemented using MPI and allows the Code to scale up 100.000 cores. The data analysis can be performed on the fly or post-processing using Python or IDL (Interactive Data Language).

A list of disk storage, number of cores and total walltime used for the simulations described in this report are shown in Table 1. The total short term storage on SuperMUC-NG for this part of the project will be 50TB. Even though most of the simulations are still ongoing, we have already achieved interesting results. Our new high-resolution run (A1) shows a different behaviour than an earlier low-resolution run (M0, [4]). Besides the two times higher resolution of A1, the difference between the two consists in the way heat conduction is modeled. To describe radiative heat transfer, we use in both runs the diffusion approximation. In Run M0, we prescribed a profile for the radiative heat conductivity, while in the PRACE project runs we use a Kramers-like opacity law, in which the heat conductivity depends on density and temperature. The latter, a more physically sound prescription, allows for the development of a layer which is convectively stable in the traditional sense, but in which the transport of heat is still upward-directed. This can have consequences for the properties of the flow but also for the magnetic field evolution [5]. We highlight one such difference in Figure 1. In the meridional cuts we show the component $\alpha_{\phi\phi}$ for run M0 (left), versus that of our new run A1 (right). The presence of negative/positive values in the northern/southern hemisphere is crucial for obtaining the equatorward propagation of the sunspot-producing zones observed during the solar cycle. In M0, the $\alpha$-effect was of the wrong sign in most of the CZ, but with the improved description of heat conduction, the region where the effect is of the correct sign has grown significantly. In the new run (A1), we now observe a reversed migration direction of the magnetic field, which shows that the changes in the $\alpha$-effect profile are significant, see Figure 2.

Table 1: Summary of the employed resources. Boldface: runs discussed in this report.

| Run | Grid size | # CPU | Total Walltime [h] | Total memory required |
|---|---|---|---|---|
| **A1** | 256x512x256 | **512** | **1440** | **17GB** |
| **A1TF** | 256x512x256 | **512** | **969** | **83GB** |
| A2 | 512x1024x512 | 2048 | 1440 | 137GB |
| A2TFNL | 512x1024x512 | 4096 | 969 | 1.52TB |
| A3 | 1024x2048x1024 | 4096 | 1440 | 1.10TB |
| A3TFNL | 1024x2048x1024 | 8192 | 969 | 12.20TB |
| A4 | 2048x4096x2048 | 8192 | 1632 | 8.90TB |

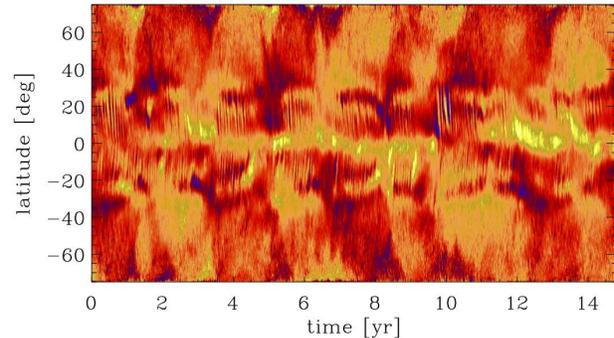

Figure 2: Time-latitude diagram from run A1, showing the azimuthally averaged azimuthal magnetic field component near the surface as function of time. In the Sun, a prominent pattern of butterfly wings, formed by the sunspot locations migrating towards the equator during the solar cycle, is visible, but reproducing this pattern has turned out to be very challenging from direct numerical simulations. Also in our new Run (A1), such a pattern is mostly absent. The physically-motivated heat conduction was seen to strongly influence it.

## Ongoing Research / Outlook

We presented here the first test-field measurements from our higher-resolution runs with improved heat conduction description. They indicate significant changes in the profiles of the most crucial inductive effect related to solar and stellar dynamo mechanisms, which already has important consequences for understanding these dynamos. Our even higher resolution runs (A2-A4) currently undertaken, will bring us into an even more turbulent regime, where magnetic background fluctuations are generated by small-scale dynamo action. To measure the turbulent effects from these runs, we will apply the novel compressible test-field method. Such analysis will allow us to study, for the first time, the interaction of small- and large-scale dynamos in a quantitative way.

## References and Links

[1] https://www.aalto.fi/en/department-of-computer-science/astroinformatics
[2] https://www.mps.mpg.de/solar-stellar-magnetic-activity
[3] http://pencil-code.nordita.org
[4] Warnecke J. et al. 2018, A&A, 609, A51.
[5] Käpylä P. et al. 2019, GAFD, 113, 149-183.

Last updated: July 2020